# Direct comparison of graphene devices before and after transfer to different substrates

Raymond Sachs, Zhisheng Lin, Patrick Odenthal, Roland Kawakami, and Jing Shi

Department of Physics and Astronomy, University of California, Riverside CA 92521

The entire graphene field-effect-transistor (FET) devices first fabricated on $SiO_2$/Si are peeled from the surface and placed on a different wafer. Both longitudinal and transverse resistivity measurements of the devices before and after the transfer are measured to calculate the mobility for a direct comparison. After transferred to different $SiO_2$/Si wafers, the mobility generally is comparable and the defect density does not show any significant increase, which indicates the degradation due to the transfer process itself is minimal. The same method can be applied to transfer graphene devices to any arbitrary substrates (e.g. $SrTiO_3$ or STO). The transfer method developed here not only eliminates the need to locate single-layer graphene on non-$SiO_2$/Si substrates for patterning, but also provides a convenient way to study the effects of various substrates on graphene electronic properties.

Mechanical exfoliation has been the very first method for attaining single layer graphene [1]. Doped Si wafers with a 300 nm thick layer of thermally grown $SiO_2$ has been used as the standard substrate because of the ease of locating via optical microscopy [2]. Due to a light interference effect on this specific thickness of $SiO_2$, single-layer graphene is distinguishable from bilayer and multilayer graphene via a color contrast [3, 4]. This makes establishing the locations of graphene flakes for patterning very efficient with a much higher throughput than locating via atomic force microscopy, scanning electron microscopy, or Raman mapping. Much work has been done to research the electronic [1, 5, 6], mechanical [7, 8], and thermal [9, 10] properties and scattering mechanisms [11-14] in graphene on $SiO_2$. There are few other substrates that have been used to exfoliate and pattern graphene devices [15-17], however comparisons of electronic properties can only be made with similar devices on $SiO_2$. Graphene can also be grown via chemical vapor deposition [18] and transferred to any arbitrary substrate [19], but again, the measurements can only be compared to the well established properties of similar devices on $SiO_2$. Therefore, it is crucial to develop a technique for measuring the same graphene flake on two different substrates for a direct comparison. In this letter, we report the transfer of entire, prefabricated graphene devices from the surface of $SiO_2$/Si to the surface of a separate $SiO_2$/Si or $SrTiO_3$ wafer. Electrical measurements are made before and after transfer to study the effect on the graphene flake and record any degrading effects. This approach is practical for investigating the properties of graphene on any arbitrary substrate without the need for locating, which proves quite difficult, or patterning with electron beam lithography which can be complicated when using thick dielectrics due to an extreme surface charging effect.

Single-layer graphene flakes were exfoliated from Kish graphite onto the surface of doped Si wafers with a 300 nm thick layer of thermally grown $SiO_2$. The locations and thicknesses of the flakes were determined using optical microscopy and Raman spectroscopy. Devices with Hall bar geometry were patterned with standard electron beam lithography techniques and metal deposition. Electrical measurements were recorded using standard AC lock-in techniques with a constant, perpendicular magnetic field of ±1500 Gauss and a back-gate applied to the doped Si. The 4-probe longitudinal voltage is measured along with the transverse voltage with the ±1500 Gauss field for an accurate calculation of the carrier mobility. The transferring of the graphene devices was accomplished using similar methods previously reported for the transfer of exfoliated graphene flakes [20]. Figure 1 summarizes the steps of the transfer process. After measuring on the initial substrate, two layers of poly(methyl methacrylate) (PMMA) are spin-coated and hard baked for 10 min at 170 °C after each coating. The wafer is placed in a beaker of 1 molar aqueous solution of NaOH for partial etching of the $SiO_2$. The $SiO_2$ is etched enough for the release of the PMMA/graphene device layer. The wafer is then placed in room temperature deionized water and physical peeling with tweezers can carefully detach the PMMA/graphene device membrane from the surface. Once it is brought to the surface, the membrane floats due to the surface tension and hydrophobic nature of PMMA. The target substrate can be brought up from underneath to slowly pull the membrane from the water. Placing the target substrate with the PMMA/graphene device layer onto a hot plate for 10 min at 50 °C helps to bake out the interfacial water layer and remove most of the extraneous wrinkles. A 30 min acetone bath at 65 °C followed by a room temperature isopropyl alcohol bath for 10 min will remove the PMMA layer. No squirt bottles or ultrasonic cleaners can be used to remove the PMMA as it may rinse away the metal and render the device immeasurable. The device can now be electrically measured again after the transfer.

The entire device with metal electrodes is transferred to the target substrate as is every graphene and bulk graphite flake that was adhered to the surface of the initial $SiO_2$/Si wafer. By transferring to another $SiO_2$/Si wafer, the graphene FET device can still be visualized with optical microscopy. It is not expected that the graphene has incurred major folds or wrinkles induced by the transfer based on AFM studies [20], although it is possible that the metal electrodes can migrate from the surface of the graphene during transfer. Fig. 2 shows the same graphene device before and after transfer. The low magnification optical images show the device at the edge of the initial wafer and placed away from the edge on the target wafer. The high magnification optical images show the clearly visible graphene flake with all electrodes still attached. This transfer method allows a direct comparison of the electrical transport properties of the same graphene and electrodes but on different $SiO_2$ substrates. The current-voltage characteristics remain linear after transfer, indicating little change to the ohmic contacts. Fig. 4(a) shows the conductivity as a function of the gate voltage measured before and after transfer. The Dirac point of the transferred device is +9 V, shifted from +30 V in the pre-transfer device. Such a variation in the position of the Dirac point is typically seen in non-transfer devices fabricated on $SiO_2$. After aligning the Dirac point, the conductivity can be compared with the relative gate voltage. In these particular devices, the effective carrier mobility calculated from the slope is comparable (8,000 vs. 6,800 $cm^2$/Vs) on the hole side, but is higher on the electron side after transfer (2,800 vs. 5,300 $cm^2$/Vs). Since we do not always see the same trend in mobility change in a number of transferred devices, we believe the device-to-device variations in mobility change reflect the variations in the local environment on the substrate such as the charged impurity distribution as well as other types of defects. In other words, the transfer process itself, e.g. etching, rinsing, and drying, does not always degrade the sample quality; therefore, the quality of the target substrate is critical.

Quantitative analysis of the σ vs. $V_g$ curve (Fig. 4a) can be used to calculate the impurity density of the graphene before and after transfer to separate $SiO_2$ substrates. Using the charged impurity model of ref. 13, $n_i$ on the electron side prior to transfer is $1.8 \times 10^{12}$ compared to $9.2 \times 10^{11}$ cm$^{-2}$ after the transfer. The change in the charged impurity density is minimal and actually decreases. The charged impurity density on the hole side increases by only 16% after transfer. These charge density variations are commonly seen in our non-transfer graphene devices on $SiO_2$ substrates, which reflects the variations in charge environment expected with amorphous $SiO_2$ substrates. Another defect analysis can be done with Raman data (inset of Fig. 4a). The ratio of the intensities of the *D* peak and the *G* peak has been used to estimate the defect density of the graphene flake [21]. Because it is the same flake, we can focus the 532 nm Raman laser (Horiba LabRam) at the approximately same position on the graphene. The calculation of the defect density (away from the edge of the flake) again is comparable before and after. Defect density $n_D$ before transferring is $2.6 \times 10^{10}$ cm$^{-2}$ and increases slightly to $7.6 \times 10^{10}$ after transferring.

The same method can be used to transfer graphene devices from $SiO_2$ to other types of substrates. For example, STO is an interesting material since it has a high dielectric constant which is desired for studying the effect of screening on electron transport properties in graphene. However, locating graphene flakes for lithography proves very difficult due to poor contrast under an optical microscope [16]. Fig. 3 shows a comparison of optical images of the same graphene device before and after transfer to a 200 μm thick STO substrate. On STO, the graphene sheet itself is hardly visible, but can be located easily from the electrodes and probed by its electrical resistance. From the carrier density measured by the Hall coefficient, we can plot the conductivity $\sigma_{xx}$ as a function of the carrier density $n$ (Fig. 4(b)). The $\sigma_{xx}$ vs. $n$ data shows that the overall carrier mobility is not much different from that in graphene/SiO$_2$. However, the screening effect can be found in the carrier density dependence of the conductivity. When it is on SiO$_2$, its conductivity is linear in carrier density over the entire density range except in the vicinity of the Dirac point where the conductivity is affected by the residual charge fluctuations, implying a density-independent mobility (~ 3,000 cm$^2$/Vs) as in typical graphene on SiO$_2$. After transferred to STO, the conductivity is no longer linear in carrier density, indicating the effect of screening by STO [22]. In fact, the calculated mobility shows an appreciable enhancement as the density is decreased from both sides except for the region very close to the Dirac point.

In summary, we have developed a transfer method and transferred entire graphene devices to two types of substrates, i.e. SiO$_2$ and STO. The sample quality is not compromised by the transfer process. The intrinsic graphene transport properties of the transferred devices can be directly compared with the pre-transfer devices. This method is applicable for transferring pre-fabricated graphene devices to any substrates, and is possible for a simultaneous transfer of a large number of devices on a wafer. Graphene can be used as a probe to measure unusual surface properties such as carrier density hysteresis caused by ferroelectric-like dipoles in STO. This application is the subject of a another publication of ours [23].

This work is supported by a NSF/NEB grant. We thank Zhiyong Wang, Wen Hua and Jen-Ru Chen for their technical assistance.

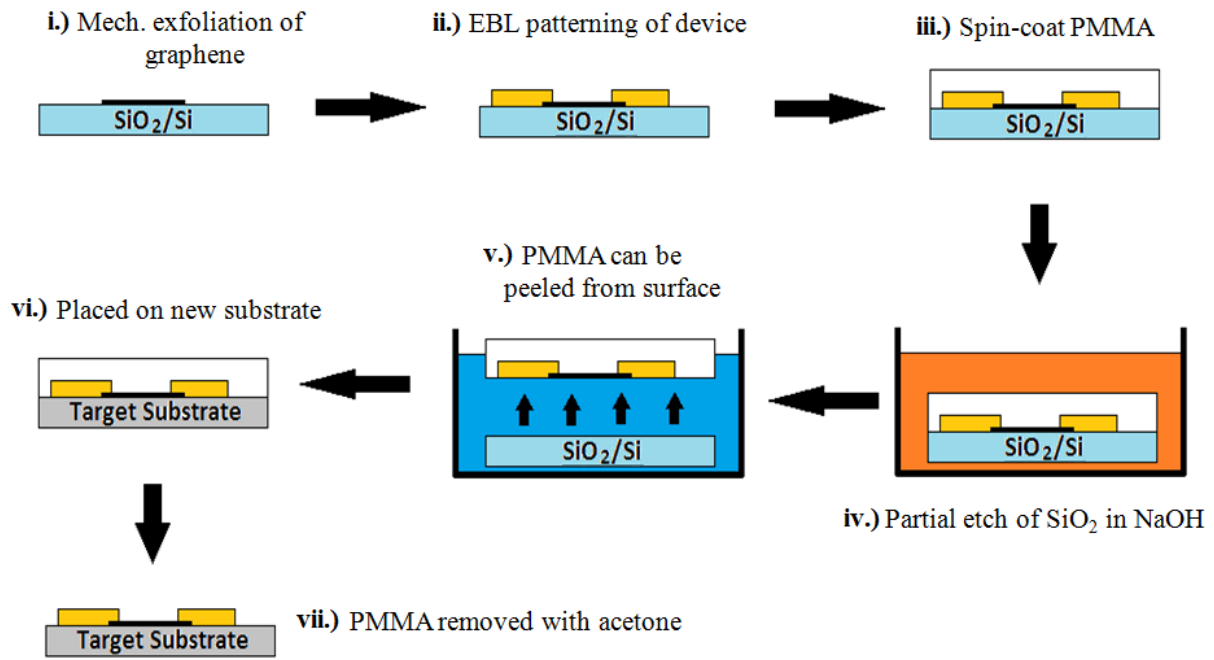

**Figure 1.** Schematic of the device transfer process. Graphene flakes are exfoliated on 300nm SiO$_2$/Si substrates and can be transferred to any target surface.

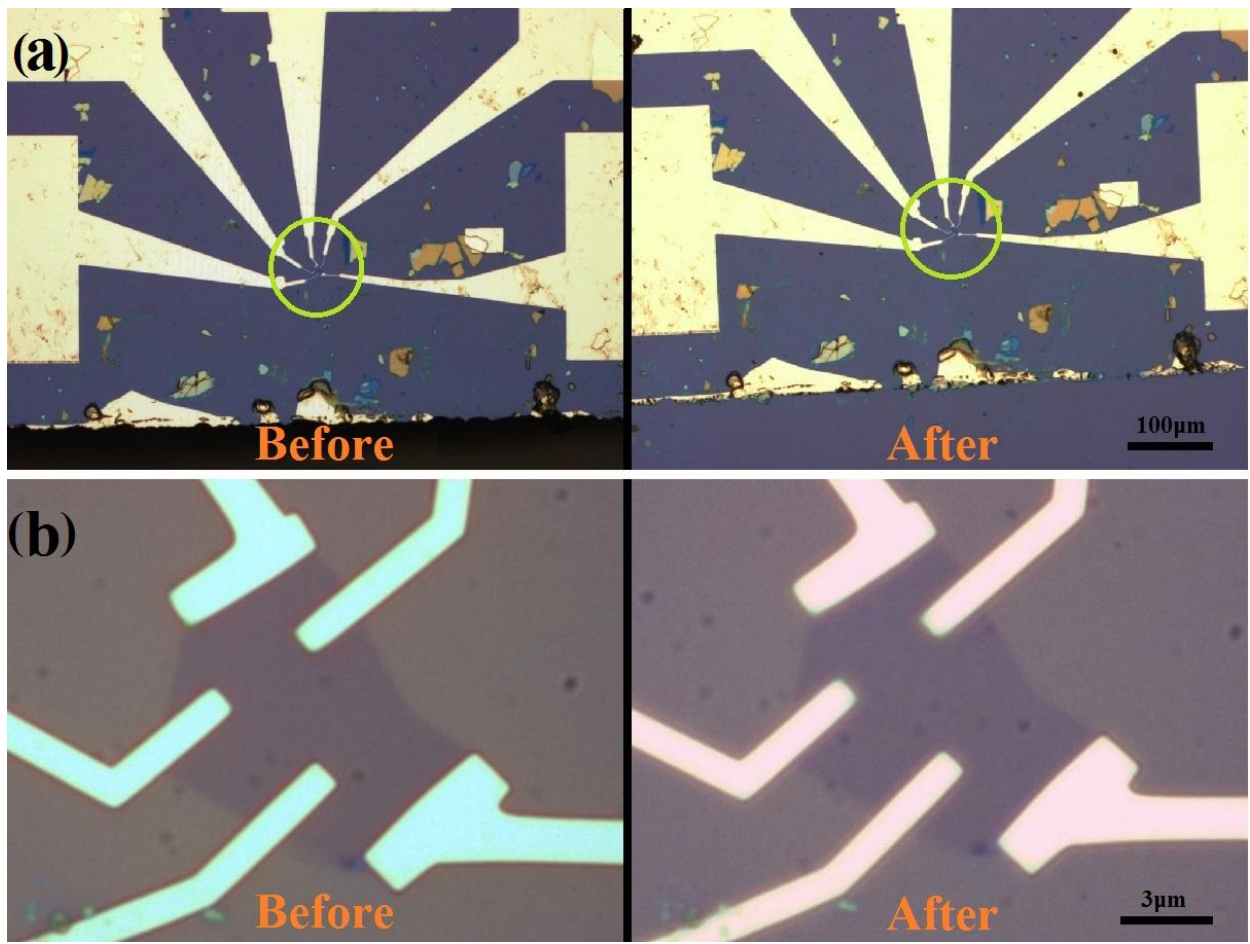

**Figure 2.** Optical images of a graphene device. Low mag. images (a) show device before and after transfer indicating it has been moved away from the edge on the target substrate. High mag. images (b) show the flake is still visible on the new SiO$_2$ surface.

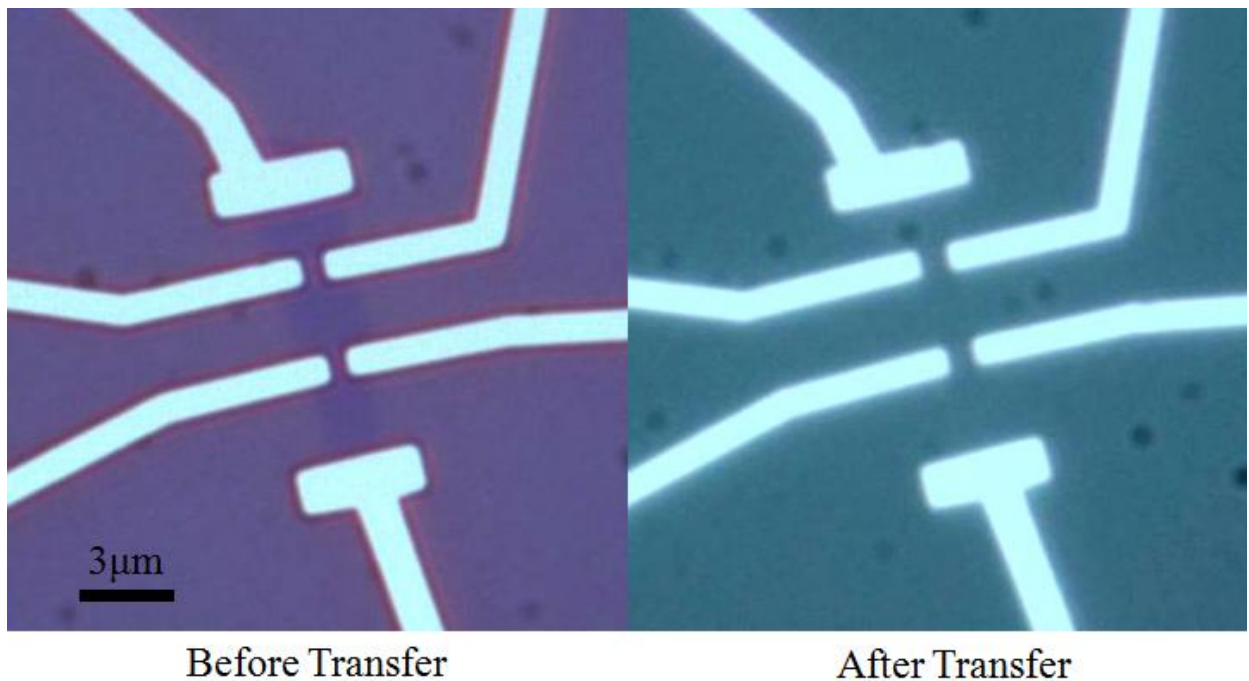

**Figure 3.** Same graphene device on SiO$_2$ (before) and on STO (after). The graphene flake on STO is hardly visible, but the device is functional.

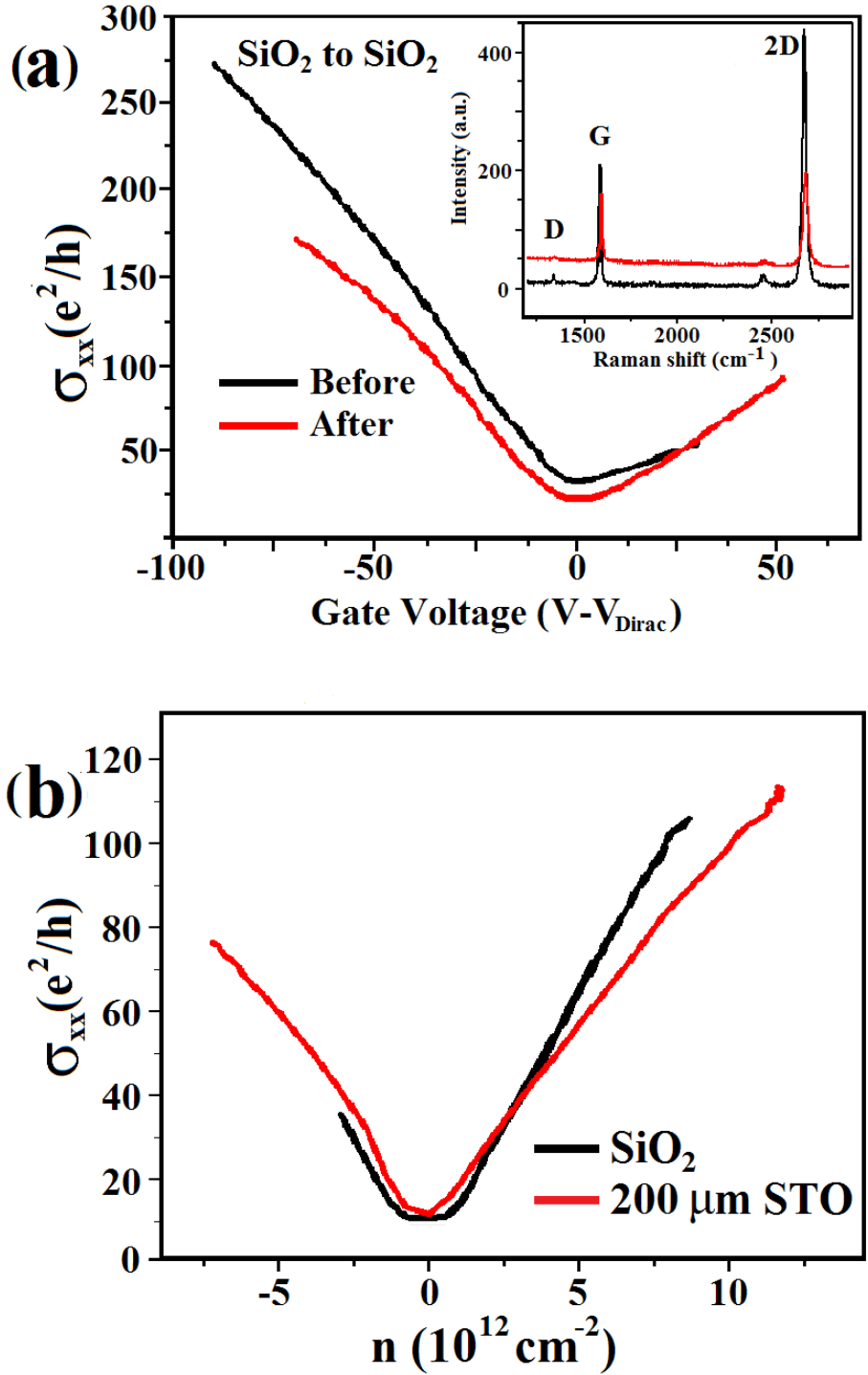

**Figure 4.** $\sigma_{xx}$ vs. $V_g$ for graphene on $SiO_2$ before and after transfer (a). Dirac points are shifted to 0 V for clarity. Differing $V_{Dirac}$ for each indicate varying charge environment. Raman spectra for graphene is used to calculate defect density from the intensity ratio $I_G/I_D$ (inset of (a)). $\sigma_{xx}$ vs. $n$ for $SiO_2$ and after transfer to STO (b). Only minor changes in mobility at high and low density are evident from the slopes of the curves.